\documentclass[12pt]{article}
\begin{document}
\title{The viscoelastic and anelastic responses of amorphous polymers in the vicinity
of the glass transition temperature}

\author{Aleksey D. Drozdov and Joachim Kaschta\\
Lehrstuhl f\"{u}r Polymerwerkstoffe\\
Institut f\"{u}r Werkstoffwissenschaften\\
Universit\"{a}t Erlangen--N\"{u}rnberg\\
Martensstrasse 7, D-91058, Erlangen, Germany}
\date{}
\maketitle

\begin{abstract}
The time-dependent response of polystyrene and poly(methyl methacrylate) is
studied in isothermal long-term shear creep tests at small strains and
various temperatures in the vicinity of the glass transition point.
A micromechanical model is derived to describe the experimental results.
Constitutive equations are developed under the assumption that 
the behavior of amorphous polymers is governed by two micro-mechanisms:
rearrangement of cooperatively relaxing regions (CRR) reflects the viscoelastic
response, whereas displacement of CRRs with respect to each other
is responsible for the anelastic response.
It is demonstrated that some critical temperature exists slightly above the
glass transition temperature, where the dependences of adjustable parameters 
on temperature are dramatically changed.
The critical temperature is associated with transition from dynamic heterogeneity
in amorphous polymers to static inhomogeneity.
\end{abstract}
\hspace*{10 mm}

\newpage

\section{Introduction}

The mechanical behavior of amorphous polymers in the vicinity of the
glass transition point, $T_{\rm g}$, has attracted essential attention in 
the past decade, see, e.g., monographs \cite{Sch90}--\cite{Don01}.
The researches focus conventionally on three main issues:
\begin{enumerate}
\item
the kinetics of slow out-of-equilibrium evolution of internal structure
for polymers equilibrated at some temperature above $T_{\rm g}$ and 
quenched to a temperature $T$ in  the sub--$T_{\rm g}$ region 
(physical aging of polymeric glasses \cite{Str78,BCK98});

\item
the existence of some critical temperatures, $T_{\rm c}$, slightly above or below
the glass transition point, associated with the crossover point predicted by the
mode-coupling theory \cite{GS92,Got99}, as well as with splitting points for 
$\alpha$ and slow $\beta$ relaxations \cite{JG70,JG72}
and for $\alpha$ and $a$ relaxations \cite{BKH98};
 
\item
temperature-induced changes in the mechanical response of amorphous polymers
from the ``glassy-like behavior" in the sub--$T_{\rm g}$ region
to the ``rubbery-like behavior" in the post--$T_{\rm g}$ domain
observed in standard creep and relaxation tests \cite{OM99}.
\end{enumerate}
The present paper is concerned with the latter two questions.

A conventional standpoint is that the entire information about the time-dependent 
behavior of solid polymers at small strains can be obtained in isothermal short-term
tests (with the duration of about 10$^{3}$ s) and the mechanical response
in creep and relaxation tests can be adequately  predicted in the framework 
of linear viscoelasticity.
To demonstrate that this is not the case, we perform long-term creep tests (with the duration
of 10$^{6}$ s) on two widely used amorphous polymers, polystyrene (PS) and poly(methyl
methacrylate) (PMMA), at various temperatures below and above $T_{\rm g}$.
Experimental data reveal rather poor superposition of creep compliance curves plotted
in double logarithmic coordinates by their shifts along the time-axis:
superposition of the middle parts of the long-term compliance curves results in deviations
of about 10 to 15 \% (in the logarithmic scale) in the regions of small and large times.
This observation implies the conclusion that viscoelasticity is not the only mechanism 
that governs the time-dependent behavior of polymers at the macro-level.
To describe another possible mechanism, the following scenario is proposed for
deformation of amorphous polymers:
\begin{enumerate}
\item
At some temperature $T_{\rm F}>T_{\rm g}$ a homogeneous state of an amorphous polymer becomes
thermodynamically unstable, and dynamic micro-heterogeneities arise in the bulk material.
These inhomogeneities (Fisher's clusters \cite{Fis93} or frustration-limited domains \cite{KZK94}) 
are observed as temporary micro-regions with higher mass densities compared to the density 
of the surrounding domains.
The presense of the Fisher clusters is confirmed experimentally in light scattering tests
which reveal long-range density fluctuations (with the lengthscale of 200 nm)
in molecular and polymeric liquids slightly above the glass transition temperature \cite{Tan00}.

\item
Below $T_{\rm F}$, an amorphous polymer is treated as an ensemble of high-density clusters
bridged by long chains (which create ``less cohesive spaces" between ``more cohesive
domains").
This picture is confirmed by low-frequency Raman spectroscopy for poly(methyl methacrylate)
\cite{ABD95} and polycarbonate \cite{MDE96} and by X-ray diffraction and 
electron microscopy for poly(ethylene terephthalate) \cite{YG67}.
Observations evidence that the less cohesive domains create a penetrating network, whereas
more cohesive regions may be thought of as isolated islands with the characteristic size
of 4 to 7 nm (below the glass transition point).
The latter result is in quantitative agreement with the characteristic lengths of cooperative
regions determined for ortho-terphenyl \cite{FDS92},
salol confined to nanopores \cite{ASG97}
and mixture of polychlorinated biphenyls \cite{RN99}
by means of dynamic light scattering, photon correlation spectroscopy and
dielectric relaxation.
The islands are separated by the less cohesive space with the average distance
between high density clusters of about 1 nm \cite{ABD95}.

\item
The domains with high density are identified with cooperatively rearranged regions (CRR)
predicted by the concept of cooperative relaxation \cite{AG65}.
Any CRR consists of dozens of strands linked by entanglements and van der Waals 
forces \cite{Sol98} and may be treated as a ``ball-like structure" with a paracrystalline-type 
order \cite{YG67}.
Rearrangements of CRRs occur at random times as they are thermally agitated.
A rearrangement event is treated as large-angle rotation of segments 
of strands in a CRR \cite{Dyr95}.
Large-angle reorientation for molecular clusters was recently confirmed 
by multidimensional deuterium nuclear magnetic resonanse spectroscopy 
for supercooled glycerol \cite{BH98} and toluene \cite{Hin98}.
Rearrangement of CRRs at the micro-level is associated with the viscoelastic response
of a specimen at the macro-level.

\item
The less cohesive space between CRRs has a complicated fractal structure.
It is modeled as an ensemble of mutually independent less cohesive domains (LCD)
with the characteristic size of about 100 nm (the length scale of long-range 
density fluctuations observed in light scattering experiments).
Two main functions are ascribed to LCDs: (i) to transmit macro-strains in a sample
to individual CRRs, and (ii) to ensure sufficient freedom for mutual displacement
of CRRs with respect to each other (within a LCD to which they belong).
Micro-deformations driven by displacements of CRRs are associated with the anelastic
response of a specimen at the macro-level.
Transformation of the internal structure of less cohesive space driven by mutual 
displacements of CRRs was confirmed by low-frequency Raman spectroscopy for 
poly(methyl methacrylate) at finite shear deformations \cite{ABD95}.
\end{enumerate}
The meaning of the term ``anelastic deformation" used in this study slightly differs 
from that conventionally employed in the analysis of the viscoelastic and viscoplastic
behavior of amorphous polymers far below their glass transition point, 
see, e.g., \cite{QPG95,PGM00}.
According to the concept of shear micro-domains \cite{Per98}, 
the anelastic deformation is treated as a deformation
which recovers during the experimental time scale at temperatures below $T_{\rm g}-20$ K, 
whereas the plastic deformation is thought of as that recoverable in the 
close vicinity of the glass transition temperature only \cite{PGM00}.
To establish some correspondence between our definition and the traditional one,
we treat the less cohesive space as a rubber-like material
and assume that some chains in the less cohevise domains are disentangled when relatively 
large stresses are applied to an amorphous polymer.
The disentanglement results in a substantial increase in the average size of a LCD
driven by coalescence of neighboring LCDs (which receive sufficient freedom to be aggregated).
This implies that the mechanical responses of a sample at loading and subsequent unloading 
do not coincide, which is evidenced as the onset of residual strains.
These strains disappear only when a sample is heated to the glass transition region
and annealed for a time sufficient for healing (reformation) of entanglements 
(which is tantamount to the return of the internal structure of a polymer to its virgin state).

It is worth noting that a similar process (disentanglement of chains in the less cohesive space
and aggregation of LCDs into larger clusters) can occur with an increase in temperature above
the glass transition point.
In this case, however, thermal fluctuations may be thought of as a driving force 
for the disentanglement process instead of mechanical stresses 
(because the intensity of fluctuations grows with temperature, $T$).
This implies some similarity between viscoplastic deformations of amorphous polymers 
below $T_{\rm g}$ and their flows at temperatures far above $T_{\rm g}$.
This similarity was recently confirmed by comparison of the activation enthalpies for 
plastic deformation of glassy polycarbonate, poly(methyl methacrylate), polystyrene,
poly(vinyl chloride) and styrene-acrylonitrile copolymer with those for the
non-Newtonian flow of their melts \cite{Nan98,NYY98}.

The objective of this study is three-fold: (i) to analyze the response of PMMA
and PS in isothermal long-term creep tests with small strains at various temperatures, 
(ii) to derive constitutive equations for the behavior of amorphous polymers 
based on the two micro-mechanisms for the time-dependent response
(rearrangement of CRRs and displacement of CRRs with respect to each other)
and (iii) to demonstrate the presence of some critical temperature, $T_{\rm c}$,
slightly above the glass transition point, where the dependences of adjustable parameters 
on temperature dramatically change.

The exposition is organized as follows.
Section 2 is concerned with the description of observations.
Constitutive equations for the time-dependent response of amorphous polymers
are developed in Section 3.
In Section 4, material constants in the stress--strain relations are found by
fitting experimental data.
Section 5 deals with a discussion of our findings.
Some concluding remarks are formulated in Section 6.

\section{Experimental}

The materials studied are commercially available polystyrene (the weight-average
molecular weight $M_{w}=352$ kg/mol and the polydispersity $M_{w}/M_{n}=2.2$)
produced by Hoechst AG 
and poly(methyl methacrylate) ($M_{w}=98$ kg/mol, $M_{w}/M_{n}=1.6$) produced by R\"{o}hm.
To determine the specific volume of polymers, $v$, as a function 
of temperature, $T$, volume dilatometric measurements were carried out 
following the procedure exposed in \cite{GS84}.
A specimen with the mass 3 g was held at the initial temperature $T_{0}=190$ $^{\circ}$C 
for 30 min to erase the effects of thermo-mechanical prehistory, and, 
afterwards, was cooled with the rate 3 K/h to the final temperature 
$T_{\ast}=-10$ $^{\circ}$C.
The error in dilatometric measurements does not exceed $10^{-4}$ cm$^{3}$/g.

The specific volumes of PS and PMMA are plotted versus temperature 
in Figures 1 and 2.
Experimental data are approximated by the linear dependence
\begin{equation}
v=v_{0}(1+\alpha_{0} T),
\end{equation}
where the adjustable parameters $v_{0}$ and $\alpha_{0}$ are found by the
least-squares technique separately in the regions of low (curve 1) and 
high (curve 2) temperatures.
The glass transition temperature, $T_{\rm g}$, is determined as a point of
intersection for straight lines 1 and 2.
The values of $T_{\rm g}$ and the coefficient
\[ \alpha=\frac{dv}{dT}=\alpha_{0}v_{0} \]
are listed in Table 1.
The results are in quantitative agreement with data provided by other
sources (see, e.g., $T_{\rm g}=100$ for PS and $T_{\rm g}=105$ $^{\circ}$C for PMMA
in \cite{Mat92} and 91.2 for PS and 98.7 $^{\circ}$C for PMMA in \cite{GS84}).
The difference between our experiments and the previous ones, see \cite{GS84}, 
consists in the following:
\begin{enumerate}
\item
in the present study polystyrene specimens were preliminary melted in vacuum 
to remove additives (paraffinic wax), whereas Greiner and Schwarzl \cite{GS84}
used them as received;

\item
different commercial grades of poly(methyl methacrylate) were tested.
\end{enumerate}

Torsional creep was measured at the stress levels that ensure
the linear response of polymers by using a high-accuracy creep tester which 
was previously described in detail in \cite{KS95}.
The samples have the shape of circular cylinders with height 105 mm and radius 6 mm.
The specimens were heated to the temperatures $T_{0}=115$ $^{\circ}$C (PS)
and $T_{0}=130$ $^{\circ}$C (PMMA) and held at this temperature  for at least
30 min to erase the effects of thermo-mechanical prehistory.
Any specimen was quenched to the test temperature, $T$, in the apparatus.
After the preconditioning time of at least 1 h, the specimen was loaded with a constant
stress, and the torsional creep compliance, $J$, was measured as a function of time $t$.
The response of PS at temperatures above 115 $^{\circ}$C was measured after heating
of samples in the apparatus to the test temperature, $T$, and preconditioning for 1 h.
The response of PMMA below the glass transition temperature was measured on specimens
which were partially equilibrated (for this purpose, the preconditioning time
was increased from 1 to 155 h depending on the test temperature).

Figures 3 and 4 show the creep compliances of PS and PMMA at various temperatures in the 
range from 95 to 140 $^{\circ}$C for PS and from 92.5 to 130 $^{\circ}$C for PMMA,
which, according to Table 1, correspond to the region of rubbery response for PS
and to the regions of glassy and rubbery behavior for PMMA.

\section{Constitutive equations}

An amorphous polymer is treated as an ensemble of cooperative rearranging
regions bridged by less cohesive space.
Denote by $X_{0}=X_{0}(T)$ the number of CRRs (per unit mass) at 
temperature $T$ and by $m=m(T)$ the average mass of a relaxing region.
The total mass of CRRs occupying unit mass of the polymer 
reads $M_{0}=mX_{0}$, and the total mass of LCDs (per unit mass) is given by
\begin{equation}
M=1-M_{0}=1-mX_{0}. 
\end{equation}
The macro-strain in a specimen, $\epsilon$, is assumed to coincide with the sum of 
the micro-strain in CRRs and LCDs, $e$,
(for simplicity, we confine ourselves to uniaxial deformation and assume 
that micro-strains are the same for all CRRs and LCDs)
and the micro-strain, $e_{\rm a}$, induced by displacement
of CRRs as rigid bodies with respect to each other,
\begin{equation}
\epsilon(t)=e(t)+e_{\rm a}(t).
\end{equation}
The subscript index ``a" indicates that mutual displacements of CRRs at the micro-level 
are associated with the anelastic response of a polymer at the macro-level.

At small strains, a LCD is modeled as a linear elastic medium
with the strain energy density (per unit mass)
\[ U_{\rm LCD}=\frac{1}{2} \mu_{\rm LCD} e^{2}, \]
where the rigidity of CLD per unit mass of a polymer, $\mu_{\rm LCD}$, 
is expressed in terms of its rigidity per unit volume, $c_{\rm LCD}$,
by the formula
\[ \mu_{\rm LCD}=c_{\rm LCD}\frac{M}{\rho}, \]
where $\rho$ is mass density of LCD (we suppose that the difference between
the densities of more cohesive regions (CRR) and less cohesive space (LCD)
is small, which implies that $\rho$ may be associated with the mass density of a polymer).

To calculate the mechanical energy of CRRs, we introduce the function
$X(t,\tau)$ which equals the number of CRRs (per unit mass of the polymer)
at time $t$ that have last been rearranged before instant $\tau$.
The function $X$ entirely determines the rearrangement process:
\begin{itemize}
\item
$X(0,0)$ coincides with the concentration of CRRs in 
a stress-free medium,
\begin{equation}
X(0,0)=X_{0},
\end{equation}

\item
the expression
\[ \frac{\partial X}{\partial \tau}(t,\tau)\biggl |_{t=\tau} d\tau \]
equals the number of CRRs (per unit mass) rearranged within the
interval $[\tau, \tau+d\tau ]$,

\item
the quantity
\[ \frac{\partial X}{\partial \tau}(t,\tau) d\tau \]
equals the number of these CRRs that have not been rearranged 
during the interval $[\tau, t]$,

\item
the amount
\[ -\frac{\partial X}{\partial t}(t,0) dt \]
equals the number of CRRs (per unit mass) that have been rearranged
for the first time during the interval $[t,t+dt]$,

\item the expression
\[ -\frac{\partial^{2} X}{\partial t\partial \tau}(t,\tau) dt\,d\tau \]
determines the number of CRRs (per unit mass) that were rearranged 
within the interval $[\tau,\tau+d\tau ]$ for the last time before
their rearrangement during the interval $[t,t+dt]$.
\end{itemize}
The viscoelastic response of a polymer is modeled
as a sequence of (driven by thermal fluctuations) random hops
of rearranging regions in their potential wells \cite{MB96}.
According to the trapping concept \cite{Dyr95,MB96}, 
any CRR is modeled in the phase space as a material point trapped 
at the bottom level of its potential well on the energy landscape.
At random times, the point hops to higher energy levels 
as the CRR is thermally agitated, but it cannot leave its trap.
Unlike previous studies, see, e.g., \cite{Dro00}, all potential wells 
are assumed to have the same depth.
This allows the number of adjustable parameters in the model
to be substantially reduced.
The account for the distribution of energies for cages does not lead to an
essential improvement in the quality of matching experimental data
and does not result in qualitative changes in our conclusions.

Referring to the transition-state theory \cite{Gol69}, we postulate
that some liquid-like (reference) energy level exists on the energy
landscape, where CRRs change their configurations.
When a CRR reaches the liquid-like state in a hop, stresses
totally relax in it.
If a relaxing region hops below the reference level,
it returns to its position at the bottom of the potential well without changes.

Let us consider a CRR at time $t$ that has last been rearranged
at instant $\tau \in [0,t)$ and denote by $e_{0}(t,\tau)$
the strain from its stress-free configuration at time $\tau$
to the deformed configuration at time $t$.
Because a CRR totally relaxes reaching the liquid-like state,
its stress-free configuration coincides with the deformed
configuration of the bulk material at the instant of rearrangement.
This implies that the strain $e_{0}(t,\tau)$ is given by
\begin{equation}
e_{0}(t,\tau)=e(t)-e(\tau).
\end{equation}
At small strains, a CRR is thought of as a linear elastic medium with 
the mechanical energy
\[ u_{\rm CRR}(t,\tau)=\frac{1}{2} c e_{0}^{2}(t,\tau),  \]
where $c$ is the rigidity per CRR.
Multiplying the mechanical energy, $u_{\rm CRR}(t,0)$,
of CRRs that have not been rearranged until time $t$ by their
number, $X(t,0)$, and using Eq. (5), we obtain the mechanical energy 
of non-rearranged CRRs (per unit mass),
\[ \frac{c}{2} X(t,0)e^{2}(t). \]
Multiplying the mechanical energy $u_{\rm CRR}(t,\tau)$, of CRRs
that have last been rearranged during the interval $[\tau,\tau+d\tau ]$
by their number $\partial X/\partial \tau(t,\tau) d\tau$, we find
the mechanical energy of these CRRs,
\[ \frac{c}{2}\frac{\partial X}{\partial \tau}(t,\tau)\Bigl [
e(t)-e(\tau) \Bigr ]^{2} d\tau. \]
Summing these expressions for various $\tau\in [0,t)$, we arrive
at the formula for the strain energy density of relaxing regions,
\[ U_{\rm CRR}(t)=\frac{c}{2} \biggl [ X(t,0)e^{2}(t)
+\int_{0}^{t} \frac{\partial X}{\partial \tau}(t,\tau)
\Bigl (e(t)-e(\tau)\Bigr )^{2} d\tau \biggr ]. \]
Neglecting the energy of mutual interaction between CRRs and between CRRs
and less cohesive space, we determine the mechanical energy 
of a polymer (per unit mass), $U$, as the sum of the strain energy 
densities for CRRs, $U_{\rm CRR}$, and LCDs, $U_{\rm LCD}$,
\begin{equation}
U(t)=\frac{1}{2} \biggl [ \frac{c_{\rm LCD}M}{\rho}
+cX(t,0)\biggr ] e^{2}(t)
+\frac{c}{2} \int_{0}^{t} \frac{\partial X}{\partial \tau}(t,\tau)
\Bigl [e(t)-e(\tau)\Bigr ]^{2} d\tau .
\end{equation}
It follows from Eqs. (3) and (6) that the function $U$ is given by
\begin{eqnarray}
U(t) &=& \frac{1}{2} \biggl [ \frac{c_{\rm LCD}M}{\rho}
+cX(t,0)\biggr ] \Bigl [\epsilon(t)-e_{\rm a}(t)\Bigr ]^{2}
\nonumber\\
&& +\frac{c}{2} \int_{0}^{t} \frac{\partial X}{\partial \tau}(t,\tau)
\Bigl [\Bigl ( \epsilon(t)-\epsilon(\tau)\Bigr )
-\Bigl ( e_{\rm a}(t)-e_{\rm a}(\tau)\Bigr ) \Bigr ]^{2} d\tau .
\end{eqnarray}
The stress, $\sigma(t)$, is expressed in terms of the macro-strain,
$\epsilon(t)$, by means of the conventional relation
\[ \sigma(t)=\rho \frac{\partial (t)}{\partial \epsilon(t)}. \]
Substitution of Eq. (7) into this formula implies that
\begin{eqnarray}
\sigma(t) &=& [ c_{\rm LCD}M +c\rho X(t,0) ] 
\Bigl [\epsilon(t)-e_{\rm a}(t)\Bigr ]
\nonumber\\      
&& +\rho c \int_{0}^{t} \frac{\partial X}{\partial \tau}(t,\tau)
\Bigl [\Bigl ( \epsilon(t)-\epsilon(\tau)\Bigr )
-\Bigl ( e_{\rm a}(t)-e_{\rm a}(\tau)\Bigr ) \Bigr ] d\tau .
\end{eqnarray}
Bearing in mind that
\[ X(t,0)+\int_{0}^{t} \frac{\partial X}{\partial \tau}(t,\tau)d\tau
= X(t,t) \]
and using Eq. (4) and the conservation law for the number of CRRs per unit mass, 
we transform Eq. (8) as follows:
\[ \sigma(t)= (c_{\rm LCD}M +c\rho X_{0})
[\epsilon(t)-e_{\rm a}(t)]
-\rho c \int_{0}^{t} \frac{\partial X}{\partial \tau}(t,\tau)
[\epsilon(\tau)-e_{\rm a}(\tau)] d\tau . \]
It follows from this formula and Eq. (2) that
\begin{equation}
\sigma(t)= G \biggl \{ [\epsilon(t)-e_{\rm a}(t)]
-\frac{b}{X_{0}} \int_{0}^{t} \frac{\partial X}{\partial \tau}(t,\tau)
[\epsilon(\tau)-e_{\rm a}(\tau)] d\tau \biggr \},
\end{equation}
where
\begin{equation}
G= c_{\rm LCD}(1-mX_{0})+c\rho X_{0},
\qquad
b=\frac{c\rho X_{0}}{c_{\rm LCD}(1-mX_{0})+c\rho X_{0}}.
\end{equation}
The constitutive equation (9) is determined by two functions, $X(t,\tau)$ 
and $e_{\rm a}(t)$, which will be determined later.

\subsection{The kinetics of rearrangement}

Denote by $\gamma$ the attempt rate (the average number of hops in a trap
per unit time) and by $Q$ the probability of reaching the liquid-like
energy level in a hop.
Multiplying the attempt rate, $\gamma$, by the probability, $Q$, 
we find the rate of rearrangement $\Gamma$,
\[ \Gamma=\gamma Q. \]
Equating $\Gamma$ to the relative rates of changes in the concentrations
of CRRs rearranged at various instants, we obtain
\begin{equation}
-\frac{1}{X(t,0)}\frac{\partial X}{\partial t}(t,0)=\Gamma,
\qquad
-\biggl [ \frac{\partial X}{\partial \tau}(t,\tau)\biggr ]^{-1}
\frac{\partial^{2} X}{\partial t \partial \tau}(t,\tau)=\Gamma .
\end{equation}
The solution of Eq. (11) with the initial condition (4) reads
\begin{equation}
X(t,0)=X_{0}\exp (-\Gamma t),\qquad
\frac{\partial X}{\partial \tau}(t,\tau)=F(\tau)
\exp [-\Gamma (t-\tau)],
\end{equation}
where
\[ F(t)=\frac{\partial X}{\partial \tau}(t,\tau)\biggl |_{\tau=t} \]
is the number of CRRs (per unit mass) rearranged per unit time at instant $t$.
To determine the value of $F(t)$, we sum the number of CRRs rearranged for
the first time at instant $t$,
\[ -\frac{\partial X}{\partial t}(t,0), \]
and the numbers of CRRs rearranged (for the last time before $t$) at some
instant $\tau\in (0,t)$ and reaching the liquid-like level at time $t$
\[ -\int_{0}^{t} \frac{\partial^{2} X}{\partial t\partial \tau}(t,\tau) 
d\tau. \]
This results in the equality
\[ F(t)=-\frac{\partial X}{\partial t}(t,0) 
-\int_{0}^{t} \frac{\partial^{2} X}{\partial t\partial \tau}(t,\tau) 
d\tau.  \]
Substitution of expressions (12) into this formula implies the linear
integral equation for the function $F(t)$,
\begin{equation}
F(t)=\Gamma\biggl [ X_{0}\exp (-\Gamma t)
+\int_{0}^{t} F(\tau) \exp \Bigl (-\Gamma (t-\tau)\Bigr ) d\tau \biggr ].
\end{equation}
The solution of Eq. (13) is given by
\[ F(t)=\Gamma X_{0}. \]
Substitution of this expression into Eq. (12) results in
\begin{equation}
\frac{\partial X}{\partial \tau}(t,\tau)=\Gamma X_{0}
\exp [-\Gamma (t-\tau)].
\end{equation}
Combining Eqs. (9) and (14), we find that
\begin{equation}
\sigma(t)=G \biggl \{ [\epsilon(t)-e_{\rm a}(t)]
-b\Gamma \int_{0}^{t} \exp \Bigl (-\Gamma (t-\tau)\Bigr )
[\epsilon(\tau)-e_{\rm a}(\tau)] d\tau \biggr \}.
\end{equation}

\subsection{The kinetics of anelastic flow}

To describe the anelastic flow (which is associated with the relative displacement 
of CRRs at the micro-level), we adopt the first order kinetic relation
\begin{equation}
\eta \frac{de_{\rm a}}{dt}=\epsilon-e_{\rm a},
\end{equation}
which may be treated as a simplifies version of the flow law
for a suspension of rigid particles in a fluid, where the left-hand side
determines the drag force, whereas the right-hand side determines
the force acting on the particles from the fluid flow.
The parameter $\eta$ stands for the drag coefficient, which, in general,
depends on the deformation history.
The simplest version of the relation between $\eta$ and $\epsilon$ 
(which should be independent of the sign of $\epsilon$) reads
\begin{equation}
\eta=\eta_{0}(1+\kappa \epsilon^{2}),
\end{equation}
where $\eta_{0}$ and $\kappa$ are material constants.
A similar equality was proposed in \cite{SC91}, where the physical meaning of
the quantities $\eta_{0}$ and $\kappa$ was discussed in terms of the 
molecular structure of polymers.
Substitution of Eq. (17) into Eq. (16) implies that
\begin{equation}
\frac{de_{\rm a}}{dt}(t)=A\frac{\epsilon(t)-e_{\rm a}(t)}
{1+\kappa \epsilon^{2}(t)},
\qquad
e_{\rm a}(0)=0,
\end{equation}
where
\[ A=\frac{1}{\eta_{0}}. \]
Given a loading program, $\epsilon=\epsilon(t)$,
Eqs. (15) and (18) entirely determine the stress $\sigma(t)$.

\subsection{Transformation of the governing equations}

Introducing the notation
\begin{equation}
\Sigma(t)=\Gamma \int_{0}^{t} \exp \Bigl (-\Gamma (t-\tau)\Bigr )
[\epsilon(\tau)-e_{\rm a}(\tau)] d\tau,
\end{equation}
we re-write Eq. (15) in the form
\begin{equation}
\sigma(t)=G \Bigl [ \epsilon (t)-e_{\rm a}(t)-b \Sigma(t)\Bigr ].
\end{equation}
It follows from Eq. (19) that the function $\Sigma(t)$ obeys the
differential equation
\begin{equation}
\frac{d\Sigma}{dt}(t)=\Gamma \Bigl [ \epsilon(t)-e_{\rm a}(t)-\Sigma(t)
\Bigr ],
\qquad
\Sigma(0)=0.
\end{equation}
Nonlinear ordinary differential equations (18) and (21) together with
the linear algebraic equation (20) describe the viscoelastic and anelastic
response of an amorphous polymer.

In the sequel, we focus on the study of creep tests with
\[ \sigma(t)=\sigma_{0} \qquad (t>0), \]
where the stress $\sigma_{0}$ ensures small deformation of a specimen.
Setting
\begin{equation}
J=\frac{\epsilon}{\sigma_{0}},
\qquad
J_{\rm a}=\frac{e_{\rm a}}{\sigma_{0}},
\qquad
S=\frac{\Sigma}{\sigma_{0}},
\qquad
J_{0}=\frac{\sigma_{0}}{G}, 
\end{equation}
we present Eq. (20) as
\begin{equation}
J(t)=J_{0}+J_{\rm a}(t)+bS(t).
\end{equation}
Substitution of expressions (22) into Eqs. (18) and (21) results in
the differential equations
\begin{eqnarray}
\frac{dJ_{\rm a}}{dt}(t) &=& A \frac{J(t)-J_{\rm a}(t)}{1+CJ^{2}(t)},
\qquad
J_{\rm a}(0)=0,
\\
\frac{dS}{dt}(t) &=& \Gamma \Bigl [ J(t)-J_{\rm a}(t)-S(t)\Bigr ],
\qquad
S(0)=0,
\end{eqnarray}
where $C=\kappa \sigma_{0}^{2}$.
Excluding the function $S$ from Eqs. (23) and (25), we arrive
at the differential equation for the function $J$,
\begin{equation}
\frac{dJ}{dt}(t)=\frac{dJ_{\rm a}}{dt}(t) +\Gamma \Bigl [ J_{0}-B\Bigl (
J(t)-J_{\rm a}(t)\Bigr )\Bigr ],
\qquad
J(0)=J_{0},
\end{equation}
where, according to Eq. (10),
\begin{equation}
B=1-b=\frac{1}{1+r},
\qquad
r=\frac{c\rho X_{0}}{c_{\rm LCD}(1-mX_{0})}.
\end{equation}
Equations (24) and (26) will be used in Section 4 for the analysis of experimental
data in creep tests.
These formulas contain 5 adjustable parameters, $A$, $B$, $C$, $J_{0}$ and $\Gamma$,
to be found by fitting observations. 
The material constants have transparent physical meaning:
$J_{0}$ is the creep compliance at the instant of application of external load,
$\Gamma$ is the characteristic rate of relaxation,
$A$ is the rate of anelastic flow at the initial interval of loading (when
an increase in $\eta$ driven by the growth of strains is negligible),
$B$ is the ratio of the stiffness of less cohesive space to the total stiffness
of a polymer,
and $C$ characterizes changes in the rate of anelastic flow induced by an increase
in strains.

\section{Validation of the model}

To find adjustable parameters in the model, we match experimental data depicted in
Figures 3 and 4.
For a creep curve at any temperature, we find the material constants 
$J_{0}$, $\Gamma$ and $B$ which minimize the discrepancies between observations 
and results of numerical simulation at the first several decades of time 
(when $\log t$ changes from $-1$ to 3 for PS and from 0 to 3 for PMMA).
The parameters $J_{0}$, $\Gamma$ and $B$ are determined by using the steepest-descent 
procedure (to simplify the analysis, $A$ is set to be zero at this stage of the analysis).
Afterwards, we fix the values of these constants found by fitting experimental data
in short-term creep tests, and determine the remaining two parameters, $A$ and $C$, which
provide the best approximation of observations when $\log t$ changes from 3 to 6.
The opportunity to divide the entire interval of measurements into two sub-intervals
(short- and long-term data) is based on the observation that the effect of anelastic
flow (which is described by the parameters $A$ and $C$) is negligible during the short-term
tests, whereas the influence of the viscoelastic response (which is described by the
constants $J_{0}$, $\Gamma$ and $B$) becomes rather small when time, $t$, essentially
exceeds the characteristic time of relaxation, $\Gamma^{-1}$.
Figures 3 and 4 demonstrate fair agreement between results of numerical simulation and
experimental data for the two amorphous polymers.

To evaluate the effect of temperature on the material constants, we plot the parameters
$\Gamma$, $A$, $B$, $C$ and $G=J_{0}^{-1}$ versus $T$.
Because the viscoelastic behavior is associated at the micro-level with rearrangement
of CRRs driven by thermal fluctuations, the dependence $\Gamma=\Gamma(T)$ is assumed to be
described by the Arrhenius law
\[ \Gamma=\Gamma_{\ast}\exp \Bigl (-\frac{\Delta E_{\Gamma}}{RT}\Bigr ), \]
where $\Gamma_{\ast}$ the rate of thermal fluctuations at high temperatures, 
$R$ is the universal gas constant
and $\Delta E_{\Gamma}$ is the energy of activation.
It follows from this equality that
\begin{equation}
\log \Gamma=\Gamma_{0}-\frac{\Gamma_{1}}{T},
\end{equation}
where 
\[ \Gamma_{0}=\log \Gamma_{\ast}, \qquad
\Gamma_{1}=\frac{\Delta E_{\Gamma}}{R}\log e,
\qquad
\log=\log_{10}. \]
Figures 5 and 6 evidence that Eq. (28) describes experimental data fairly well
when the adjustable parameters $\Gamma_{i}$ are determined independently in three
different regions:
\begin{enumerate}
\item
the low-temperature region below the glass transition point;

\item
the medium-temperature region between the glass transition temperature,
$T_{\rm g}$, and some critical temperature, $T_{\rm c}$;

\item
the high-temperature region where temperature, $T$, exceeds $T_{\rm c}$.
\end{enumerate}
The values of $T_{\rm g}$ and $T_{\rm c}$ found by matching observations are collected
in Table 2.

To compare our results with those available in the literature, we calculate the fragility
index $\bar{m}$, see \cite{BNA93}, which, in our notation, is defined as
\[ \bar{m}=-\frac{d\ln \Gamma}{d(T_{\rm g}/T)}\biggl |_{T=T_{\rm g}}. \]
It follows from this equality and Eq. (28) that
\begin{equation}
\bar{m}=\frac{\Gamma_{1}}{T_{\rm g}}\ln 10.
\end{equation}
We calculate $\bar{m}$ for the two polymers by using Eq. (29) and the data depicted in Figures 
5 and 6 and find that $\bar{m}=121.2$ for PS and $\bar{m}=98.9$ for PMMA.
These quantities provide excellent agreement with the data collected in \cite{HM01}:
$\bar{m}=116$ for PS and $\bar{m}=103$ for PMMA.

The anelastic flow of CRRs is assumed to be thermally activated, which implies
that the dependence $A=A(T)$ may also be approximated by the Arrhenius law with some
characteristic rate $A_{\ast}$ and activation energy $\Delta E_{A}$.
By analogy with Eq. (28), we write
\begin{equation}
\log A=A_{0}-\frac{A_{1}}{T}
\end{equation}
with
\[ A_{0}=\log A_{\ast}, \qquad
A_{1}=\frac{\Delta E_{A}}{R}\log e.  \]
Experimental data for $A(T)$ are depicted in Figures 7 and 8 which show that
Eq. (30) provides an acceptable approximation of observations.

The adjustable parameters $B$, $C$ and $G_{0}$ are plotted versus temperature $T$
in Figures 9 to 14.
The effect of temperature is fairly well described by the
phenomenological relations
\begin{equation}
\log B=B_{0}+B_{1}\Delta T,
\qquad
\log C=C_{0}+C_{1}\Delta T,
\qquad
\log G=G_{0}-G_{1}\Delta T,
\end{equation}
where $\Delta T=T-T_{\rm g}$ and the constants $B_{i}$, $C_{i}$ and $G_{i}$ 
are determined by the least-squares technique.
The quantity $\Delta T$, where $T_{\rm g}$ is found from the dilatometric tests,
characterizes a measure of undercooling for a polymer.
It is worth noting that ``linear" Eqs. (31) are purely phenomenological and they
cannot adequately describe the influence of temperature on the adjustable parameters
in the close vicinities of critical points, $T_{\rm g}$ and $T_{\rm c}$, where they result
in corner points or even jumps of the graphs, whereas we suppose that the functions $B(T)$,
$C(T)$ and $G(T)$ are rather smooth.

\section{Discussion}

Figures 5 and 6 demonstrate that the rate of increase, $\Gamma_{1}$, in the relaxation rate,
$\Gamma$, with temperature reaches its maximum in the transition region 
$(T_{\rm g}, T_{\rm c})$, whereas in the sub-$T_{\rm g}$ and post-$T_{\rm c}$ 
domains this rate is essentially smaller.
In other words, a decrease of temperature in the interval $(T_{\rm g}, T_{\rm c})$
by 1 K results in slowing down of the relaxation process which substantially exceeds
that for temperatures out of this interval.
This implies that changes in the internal structure of amorphous polymers 
between the glass transition point, $T_{\rm g}$, and the critical point, $T_{\rm c}$, 
are strongly affected by some additional micro-mechanism which does not ``work" 
above the critical temperature and below the glass transition temperature.
Bearing im mind that an amorphous polymer reveals a pattern of {\em dynamic heterogeneity}
at evelated temperatures, whereas it exhibits a pattern of {\em spatial inhomogeneity}
at low temperatures, the region $(T_{\rm g}, T_{\rm c})$ may be identified
as a transition region from dynamic to static heterogeneity.
Assuming this transition to be associated with the onset of surface energy of CRRs,
we may conclude that an increase in the activation energy of amorphous polymers
in the transition domain is driven by the growth of the surface energy of cooperatively 
relaxing regions.

Figures 7 and 8 show that the rate of anelastic flow, $A$, increases with temperature, $T$,
in agreement with the theory of thermally activated processes.
However, in contrast with Figures 5 and 6, these figures reveal different effects of temperature
on the parameter $A$: the rate of increase in $A$ is maximal in the transition
region $(T_{\rm g}, T_{\rm c})$ for PS, whereas the maximum of this quantity is reached in
the post-$T_{\rm c}$ region for PMMA.
Because the Fisher temperatures, $T_{\rm F}$, for these two polymers have not yet 
been determined experimentally, one can only speculate about an explanation for 
this observation.
As two possible mechanisms for this phenomenon, we would mention 
(i) the effect of side groups in poly(methyl methacrylate) molecules,
and (ii) the influence of parallelization of aromatic rings in polystyrene.

Because the mobility of side groups in PMMA is rather high in the region 
$(T_{\rm g}, T_{\rm c})$, they may create temporary junctions 
which strongly restrict mutual displacement of CRRs and slow down anelastic flow.
The onset of temporary junctions is assumed to be driven by 
van der Waals forces and dipole interactions between side groups belonging to 
different chains.
With the growth of temperature, $T$, the strength of these junctions decreases,
whereas the intensity of thermal fluctuations increases, which implies that
above the critical temperature, $T_{\rm c}$, the rate of anelastic flow in PMMA grows
more rapidly, since the influence of temporary crosslinks between side groups becomes 
negligible (curves 2 and 3 in Figure 8). 

The decrease in the slope of the graph $A(T)$ for PS at temperatures above $T_{\rm c}$
may be associated with parallelization of aromatic rings belonging to different macromolecules
(inter-chain interactions) which implies (partial) molecular ordering 
and a decrease in the segmental mobility in the less cohesive space \cite{BSA93}.
Both these factors result in a decay in the rate of mutual displacement of CRRs.
Below the critical temperature parallelization does not occur because the energy 
of thermal fluctuations is insufficient for an appropriate ordering of aromatic rings
(curves 1 and 2 in Figure 7).

Figures 9 and 10 demonstrate that the parameter $B$ decreases below the critical temperature,
$T_{\rm c}$, reaches its minimum at $T_{\rm c}$, and afterwards, monotonically increases
with temperature.
It follows from Eq. (27) that $B$ is inversely proportional to $r$,
where $r$ is the ratio of the rigidity of CRRs to the rigidity of LCDs.
The increase in $B$ (which is tantamount to a decrease in $r$) with temperature 
at $T>T_{\rm c}$ is quite natural, because the rigidity of less cohesive space grows
(according to the theory of rubber elasticity, the elastic modulus is proportional
to temperature), whereas the rigidity of CRRs decreases and vanishes at the Fisher
temperature, $T_{\rm F}$.
The decrease in $B$ (or the increase in $r$) with temperature, $T$,
below $T_{\rm c}$ means that under cooling of a sample, the rigidity 
of less cohesive space between CRRs grows faster than the rigidity of 
relaxing regions.
This may be explained by the growth of the surface energy for CRRs,
which substantially reduces mobility of strands that bridge rearranging domains
and enhances their adsorption on the boundaries of CRRs,
which, in turn, leads to an increase in the mechanical energy of chains 
in the less cohesive domains \cite{HV98}.
This explanation is supported by the fact that temperature-induced changes in
the parameter $B$ become substantially weaker in the sub-$T_{\rm g}$ region
(Figure 10). 

According to Figures 11 and 12, the parameter $C$ decreases with temperature
in the sub-$T_{\rm g}$ region and increases in the post-$T_{\rm g}$ region.
The decrease in $C$ with temperature, $T$, below the glass transition point 
seems plausible, because it is equivalent to an increase in the rate of anelastic 
flow, see Eq. (24).
The increase in $C$ with temperature in the rubbery domain ($T>T_{\rm g}$)
is not rather pronounced (Figures 11 and 12 present semi-logarithmic plots of the data), 
and it may be explained by the onset of the dynamic heterogeneity in polymers: due to thermal
fluctuations, CRRs disappear earlier than they perform substantial displacements 
with respect to one another.
This conclusion is confirmed by results depicted in Figure 11 which demonstrate
that thermally induced changes in $C$ practically vanish above the critical temperature,
$T_{\rm c}$, where dynamic heterogeneity plays the key role.

Figures 13 and 14 show that the critical temperature, $T_{\rm c}$, cannot be identified
from the dependence of the elastic modulus, $G$, on temperature: the parameter $G$ 
monotonically decreases with $T$ in the entire post-$T_{\rm g}$ region.
According to Figure 14, the rate of decrease in $G$ is higher in the post-$T_{\rm g}$
domain than in the sub-$T_{\rm g}$ region.
The latter seems physically plausible if we suppose that changes in $G$ in the post-$T_{\rm g}$
region are affected by two processes: (i) thermally-driven decrease in the rigidity, $c$,
and (ii) reduction of the surface energy for CRRs.
A weak effect of the critical temperature on elastic moduli may be a reason 
why the effects studied in this work have not been previously revealed in conventional
short-term tests.

The temperatures, $T_{\rm g}$ and $T_{\rm c}$, found as critical points for the dependences
of material parameters on temperature, $T$, are collected in Table 2.
The table demonstrates rather small scatters of data, which may serve as an indirect 
validation of the model. 
In particular, it is found that the average glass transition temperature for PMMA
found by matching observations in the mechanical tests is extremely close to that 
determined in the dilatometric tests (Tables 1 and 2).
The latter conclusion is not surprising because the rate of cooling in dilatometric
tests is comparable with the creep rate in long-term mechanical experiments.

The critical temperatures for the two polymers are rather close to each other,
which is in agreement with observations which evidence that PS and PMMA have similar 
values of most physical parameters \cite{Kre90}.
The latter conclusion is confirmed by the data listed in Table 3
which show quantitatively the same effects of temperature on adjustable parameters 
for PS and PMMA (the constants which appear to be similar are printed in bold).
The only exception from this rule is the rate of anelastic flow, $A$, which seems to
depend strongly on the molecular weight and chemical structure of the polymers.

\section{Concluding remarks}

Experimental data are presented in long-term shear creep tests on polystyrene
and poly\-(me\-thyl methacrylate) at various temperatures near the glass transition
points.
Constitutive equations are developed for the uniaxial mechanical response of amorphous
polymers at small strains.
The model is based on the hypothesis that the time-dependent behavior of polymers 
is governed by two micro-mechanisms: rearrangement of CRRs describes 
the viscoelasticity, whereas mutual displacement of CRRs reflects the anelastic phenomena.
Adjustable parameters in the stress--strain relations are found as functions of
temperature.
It is revealed that some critical temperature exists for the polymers, $T_{\rm c}\approx
T_{\rm g}+20$ K, where the dependences of material constants on temperature
substantially change.

\subsubsection*{Acknowledgement}

We would like to express our gratitude to Prof. H. M\"{u}nstedt
for numerous discussions of the subject and for his helpful comments
on this work.

\newpage

\section*{List of captions}

\hspace{6 mm}
{\bf Figure 1:} The specific volume $v$ cm$^{3}$/g versus temperature $T$ $^{\circ}$C
for polystyrene.
Circles: observations.
Solid lines: approximation of the experimental data by Eq. (1).
Curve 1: $v_{0}=0.9529$, $\alpha_{0}=2.2213\times 10^{-4}$ K$^{-1}$;
curve 2: $v_{0}=0.9173$, $\alpha_{0}=6.4033\times 10^{-4}$ K$^{-1}$

{\bf Figure 2:} The specific volume $v$ cm$^{3}$/g versus temperature $T$ $^{\circ}$C
for poly(methyl methacrylate).
Circles: observations.
Solid lines: approximation of the experimental data by Eq. (1).
Curve 1: $v_{0}=0.8332$, $\alpha_{0}=2.6172\times 10^{-4}$ K$^{-1}$;
curve 2: $v_{0}=0.7965$, $\alpha_{0}=7.3739\times 10^{-4}$ K$^{-1}$

{\bf Figure 3:} The creep compliance $J$ GPa$^{-1}$ versus time $t$ s
for polystyrene at a temperature $T$ $^{\circ}$C.
Circles: experimental data.
Solid lines: results of numerical simulation.
Curve 1: $T=95.0$;
curve 2: $T=97.5$;
curve 3: $T=100.0$;
curve 4: $T=105.0$;
curve 5: $T=110.0$;
curve 6: $T=115.0$;
curve 7: $T=120.0$;
curve 8: $T=130.0$;
curve 9: $T=140.0$

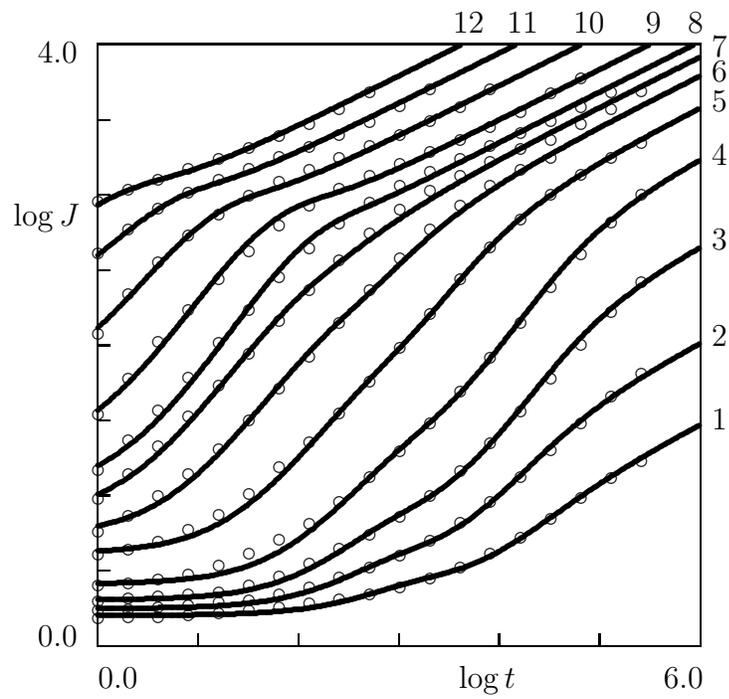
{\bf Figure 4:} The creep complience $J$ GPa$^{-1}$ versus time $t$ s
for poly(methyl methacrylate) at a temperature $T$ $^{\circ}$C.
Circles: experimental data.
Solid lines: results of numerical simulation.
Curve 1: $T=92.5$;
curve 2: $T=97.5$;
curve 3: $T=100.0$;
curve 4: $T=102.5$;
curve 5: $T=105.0$;
curve 6: $T=107.5$;
curve 7: $T=110.0$;
curve 8: $T=112.0$;
curve 9: $T=115.0$;
curve 10: $T=120.0$;
curve 11: $T=125.0$;
curve 12: $T=130.0$

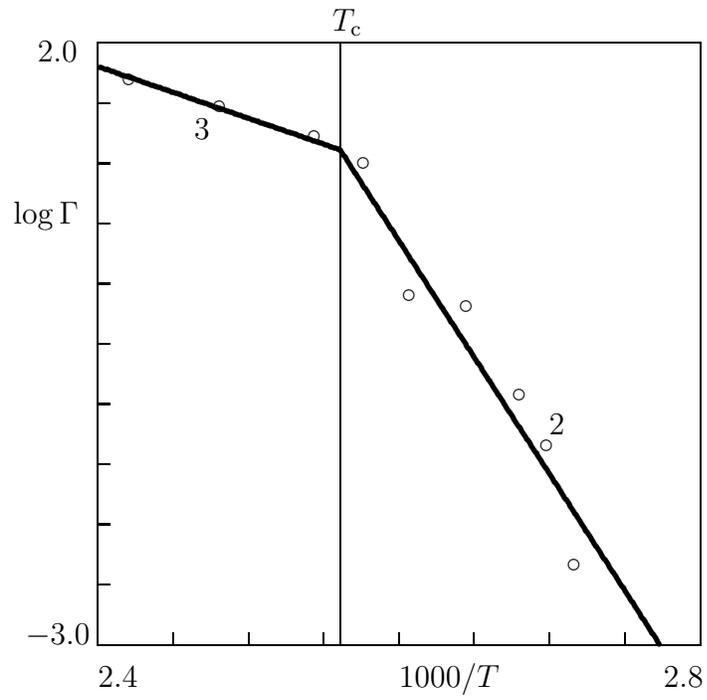
{\bf Figure 5:} The relaxation rate $\Gamma$ s$^{-1}$ versus temperature 
$T$ K for polystyrene.
Circles: treatment of observations.
Solid lines: approximation of the experimental data by Eq. (28).
Curve 2: medium-temperature region, $\Gamma_{0}=50.73$, $\Gamma_{1}=19373.5$;
curve 3: high-temperature region, $\Gamma_{0}=12.12$, $\Gamma_{1}=4298.9$

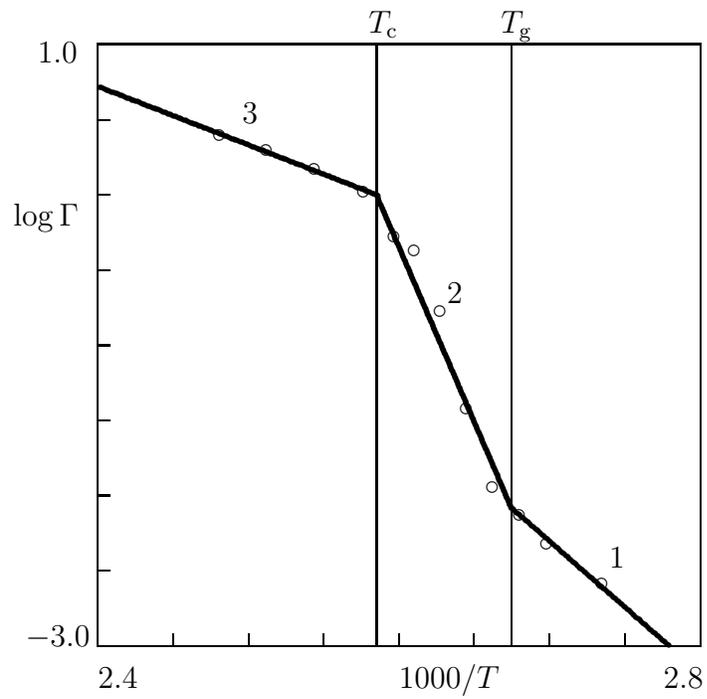
{\bf Figure 6:} The relaxation rate $\Gamma$ s$^{-1}$ versus temperature 
$T$ K for poly(methyl metha\-cry\-late).
Circles: treatment of observations.
Solid lines: approximation of the experimental data by Eq. (28).
Curve 1: low-temperature region, $\Gamma_{0}=21.35$, $\Gamma_{1}=8759.6$;
curve 2: medium-temperature region, $\Gamma_{0}=60.10$, $\Gamma_{1}=23245.7$;
curve 3: high-temperature region, $\Gamma_{0}=10.07$, $\Gamma_{1}=3893.6$

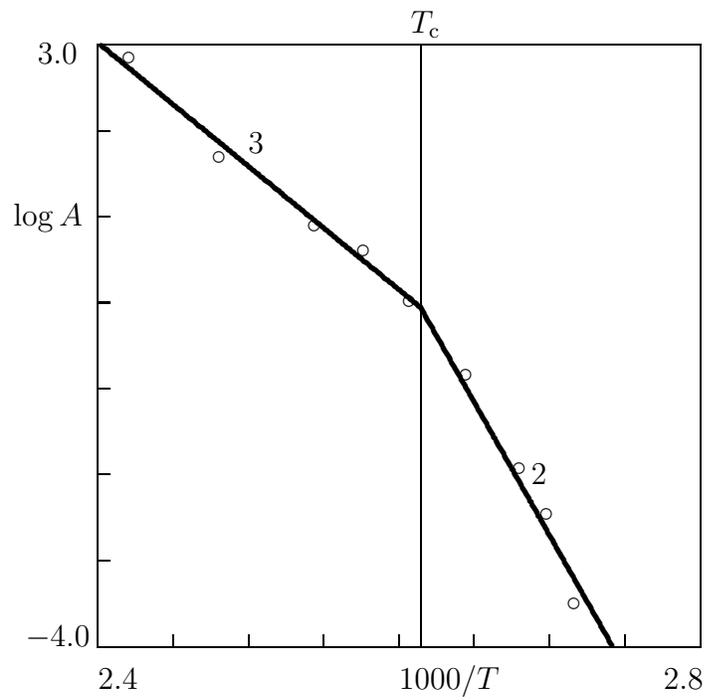
{\bf Figure 7:} The rate of anelastic flow $A$ s$^{-1}$ versus temperature 
$T$ K for polystyrene.
Circles: treatment of observations.
Solid lines: approximation of the experimental data by Eq. (30).
Curve 2: medium-temperature region, $A_{0}=80.79$, $A_{1}=30921.1$;
curve 3: high-temperature region, $A_{0}=37.52$, $A_{1}=14371.1$

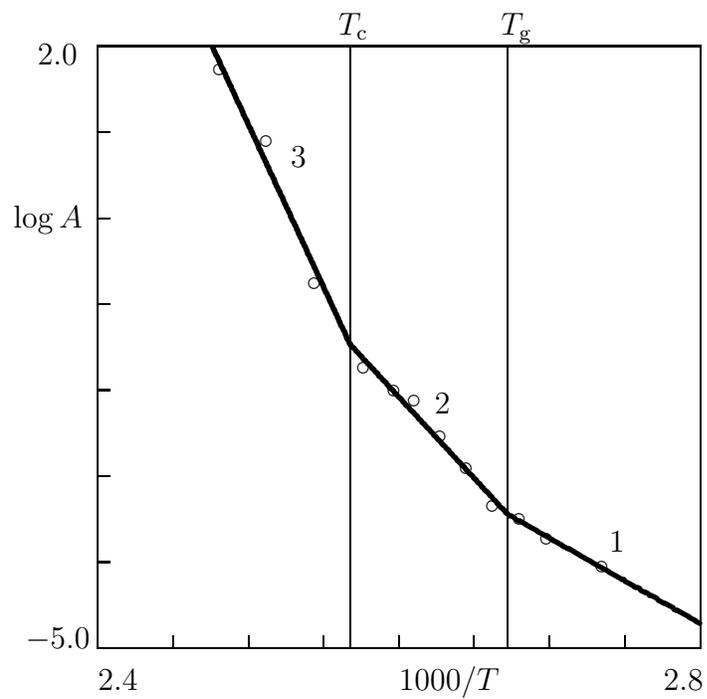
{\bf Figure 8:} The rate of anelastic flow $A$ s$^{-1}$ versus temperature 
$T$ K for poly(methyl methacrylate).
Circles: treatment of observations.
Solid lines: approximation of the experimental data by Eq. (30).
Curve 1: low-temperature region, $A_{0}=23.11$, $A_{1}=9934.8$;
curve 2: medium-temperature region, $A_{0}=47.03$, $A_{1}=18886.4$;
curve 3: high-temperature region, $A_{0}=95.73$, $A_{1}=37851.8$

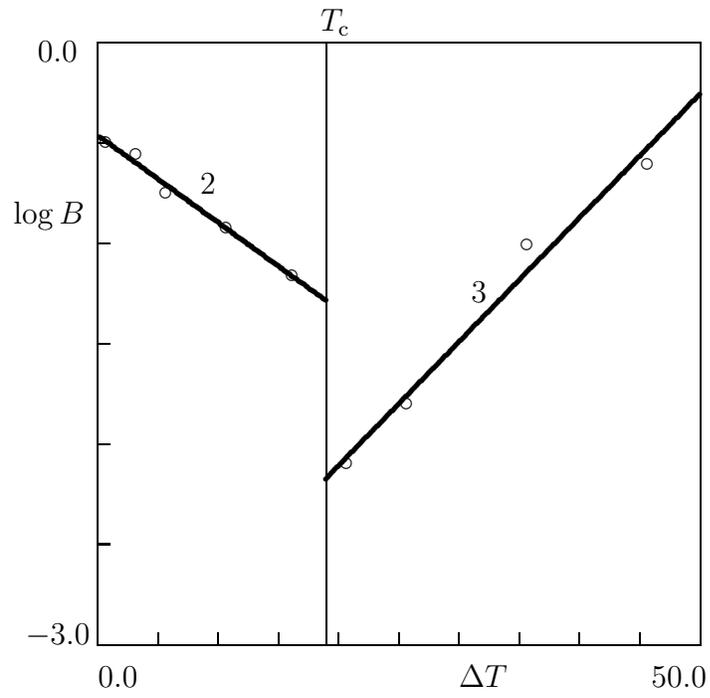
{\bf Figure 9:} The dimensionless parameter $B$ versus the degree of undercooling
$\Delta T$ K for polystyrene.
Circles: treatment of observations.
Solid lines: approximation of the experimental data by Eq. (31).
Curve 2: medium-temperature region, $B_{0}=-0.4596$, $B_{1}=-0.0435$;
curve 3: high-temperature region, $B_{0}=-3.3427$, $B_{1}=0.0617$

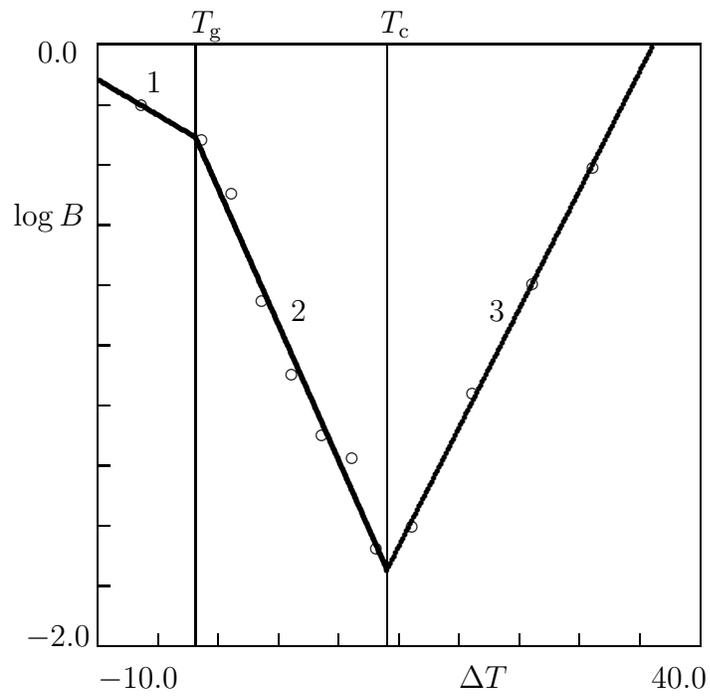
{\bf Figure 10:} The dimensionless parameter $B$ versus the degree of undercooling
$\Delta T$ K for poly(methyl methacrylate).
Circles: treatment of observations.
Solid lines: approximation of the experimental data by Eq. (31).
Curve 1: low-temperature region, $B_{0}=-0.3521$, $B_{1}=-0.0236$;
curve 2: medium-temperature region, $B_{0}=-0.4778$, $B_{1}=-0.0904$;
curve 3: high-temperature region, $B_{0}=-2.8531$, $B_{1}=0.0789$

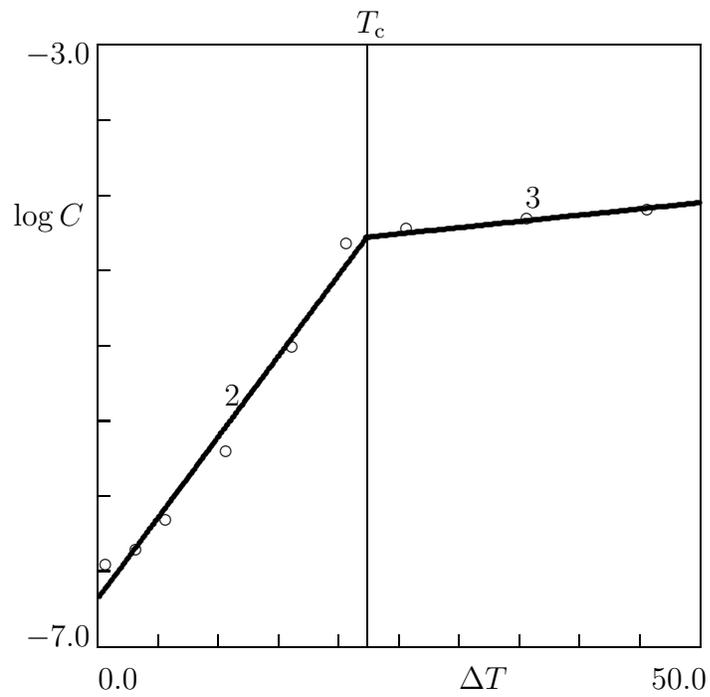
{\bf Figure 11:} The parameter $C$ GPa$^{2}$ versus the degree of undercooling
$\Delta T$ K for polystyrene.
Circles: treatment of observations.
Solid lines: approximation of the experimental data by Eq. (31).
Curve 2: medium-temperature region, $C_{0}=-6.6820$, $C_{1}=0.1075$;
curve 3: high-temperature region, $C_{0}=-4.4625$, $C_{1}=0.0083$

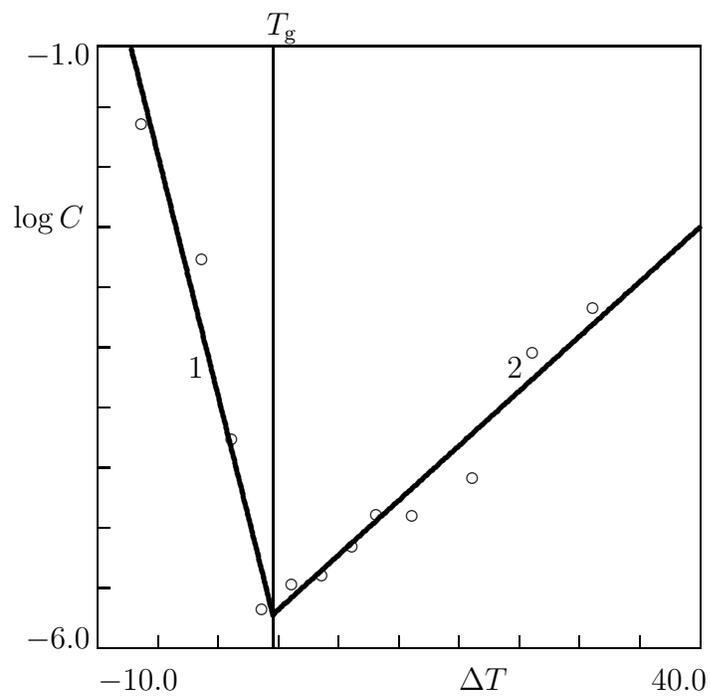
{\bf Figure 12:} The parameter $C$ GPa$^{2}$ versus the degree of undercooling
$\Delta T$ K for poly(methyl methacrylate).
Circles: treatment of observations.
Solid lines: approximation of the experimental data by Eq. (31).
Curve 1: glassy region, $C_{0}=-3.9026$, $C_{1}=-0.4004$;
curve 2: rubbery region, $C_{0}=-6.1357$, $C_{1}=0.0908$

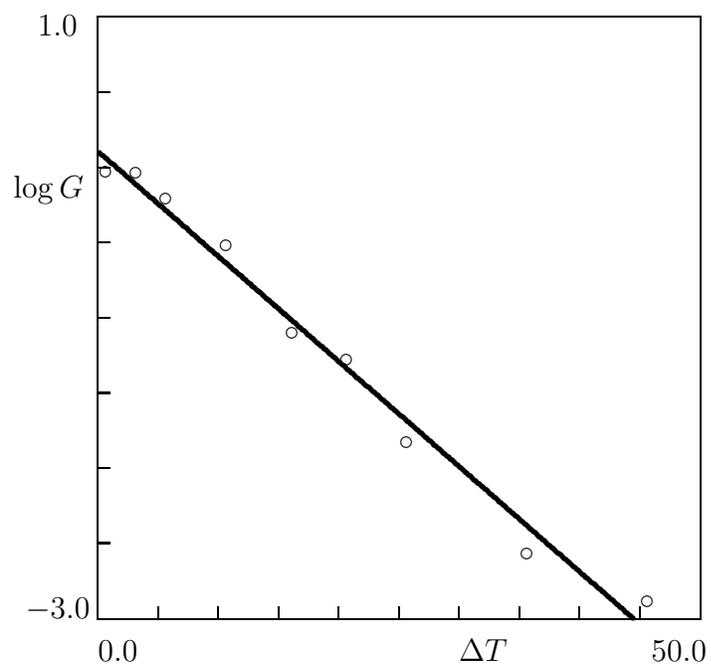
{\bf Figure 13:} The elastic modulus $G$ GPa versus the degree of undercooling
$\Delta T$ K for polystyrene.
Circles: treatment of observations.
Solid line: approximation of the experimental data by Eq. (31)
with $G_{0}=0.1094$ and $G_{1}=0.0699$

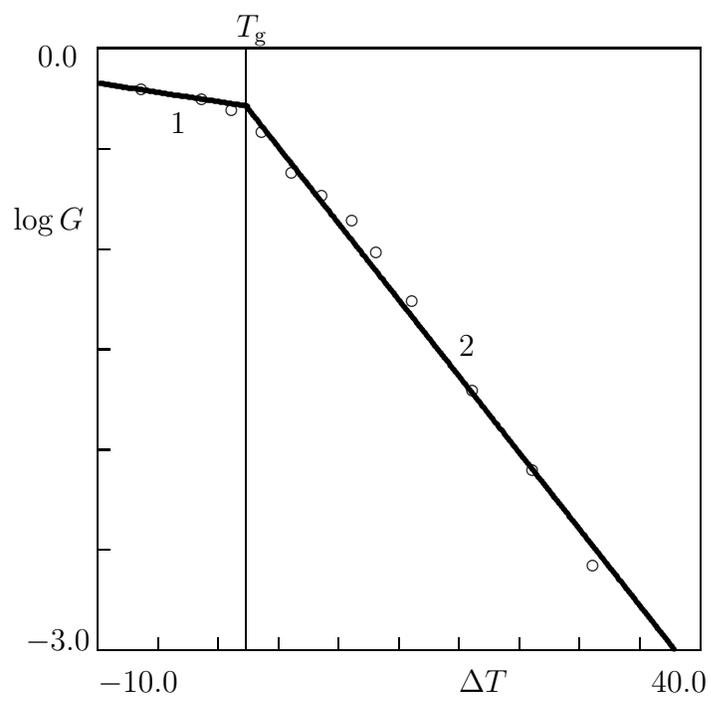
{\bf Figure 14:} The elastic modulus $G$ GPa versus the degree of undercooling
$\Delta T$ K for poly(methyl methacrylate).
Circles: treatment of observations.
Solid lines: approximation of the experimental data by Eq. (31).
Curve 1: glassy region, $G_{0}=-0.2635$, $G_{1}=0.0093$;
curve 2: rubbery region, $G_{0}=-0.1092$, $G_{1}=0.0762$
\newpage

\noindent
Table 1: Material constants determined by dilatometric measurements
\vspace*{6 mm}

\begin{center}
\begin{tabular}{@{} l c c c @{}}
\hline
Polymer & $T_{\rm g}$ $^{\circ}$C & $\alpha_{\rm glass}\times 10^{4}$ $K^{-1}$ & 
$\alpha_{\rm rubber}\times 10^{4}$ $K^{-1}$ \cr
\hline
PS   & 94.8 & 2.12 & 5.87 \\
PMMA & 99.5 & 2.18 & 5.87 \\
\hline
\end{tabular}
\end{center}
\vspace*{10 mm}

\noindent
Table 2: The temperatures $T_{\rm c}$ and $T_{\rm g}$ determined as critical points 
for the dependences of adjustable parameters on temperature
\vspace*{6 mm}

\begin{center}
\begin{tabular}{@{} c c c c @{}}
\hline
Parameter & $T_{\rm c}$ $^{\circ}$C (PS) & $T_{\rm c}$ $^{\circ}$C (PMMA) & 
$T_{\rm g}$ $^{\circ}$C (PMMA) \cr
\hline
$\Gamma$ & 117.3 & 113.6 & 100.7 \\
$A$      & 109.3 & 116.3 & 101.0 \\
$B$      & 113.4 & 112.9 &  97.0 \\
$C$      & 116.8 & ---    & 103.5 \\
$G_{0}$  & ---    & ---    & 101.2 \cr
\hline
Average  & 114.2 & 114.3 & 100.7 \\
\hline
\end{tabular}
\end{center}
\vspace*{10 mm}

\noindent
Table 3: Material constants determined by mechanical measurements
\vspace*{6 mm}

\begin{center}
%
\end{center}
\vspace*{10 mm}

\caption{}
\end{figure}

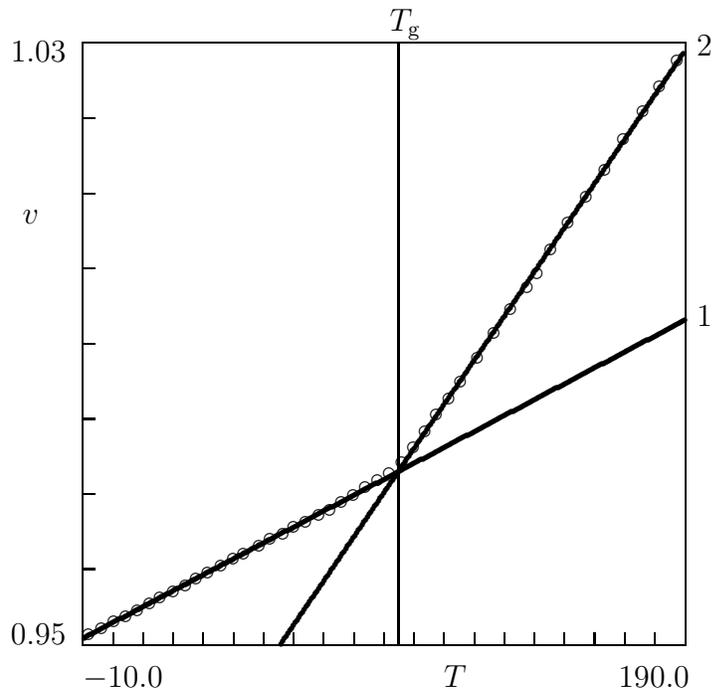
\begin{figure}[t]
\begin{center}
%
\end{center}
\vspace*{10 mm}

\caption{}
\end{figure}

\begin{figure}[t]
\begin{center}
%
\end{center}
\vspace*{10 mm}

\caption{}
\end{figure}

\begin{figure}[t]
\begin{center}
%
\end{center}
\vspace*{10 mm}

\caption{}
\end{figure}

\end{document}